\documentclass[journal=nalefd,manuscript=letter,layout=twocolumn]{achemso}
\setkeys{acs}{etalmode=truncate,maxauthors=10,articletitle=true} 
\usepackage{graphicx}
\usepackage{dcolumn}
\usepackage{bm}
\usepackage{gensymb}
\usepackage{siunitx}
\usepackage{amsmath}
\usepackage{amssymb}

\newcommand{\dg}[1]{\ensuremath{#1^\circ}}
\newcommand{\wavenumber}[1]{\ensuremath{\SI{#1}{\per\centi\meter}}}
\newcommand{\musec}[1]{\ensuremath{\SI{#1}{\micro\second}}}
\newcommand{\muJ}[1]{\ensuremath{\SI{#1}{\micro\joule}}}
\newcommand{\mumetr}[1]{\ensuremath{\SI{#1}{\micro\meter}}}
\newcommand{\nmetr}[1]{\ensuremath{\SI{#1}{\nano\meter}}}

\newcommand{\FigureFormat}{pdf}

\author{Nikolai Christian Passler}
 \email{passler@fhi-berlin.mpg.de}
 \affiliation{Fritz Haber Institute of the Max Planck Society, Faradayweg 4-6, 14195 Berlin, Germany}
\author{Andreas He\ss ler}
 \affiliation{Institute of Physics (IA), RWTH Aachen University, 52056 Aachen, Germany}
\author{Matthias Wuttig}
 \affiliation{Institute of Physics (IA), RWTH Aachen University, 52056 Aachen, Germany}
\author{Thomas Taubner}
 \affiliation{Institute of Physics (IA), RWTH Aachen University, 52056 Aachen, Germany}
\author{Alexander Paarmann}
 \email{alexander.paarmann@fhi-berlin.mpg.de}
 \affiliation{Fritz Haber Institute of the Max Planck Society, Faradayweg 4-6, 14195 Berlin, Germany}

\title{Surface Polariton-Like s-Polarized Waveguide Modes in Switchable Dielectric Thin-Films on Polar Crystals}


\begin{document}

%
%
%

\begin{abstract}
Surface phonon polaritons (SPhP) and surface plasmon polaritons (SPP), evanescent modes supported by media with negative permittivity, are a fundamental building block of nanophotonics. These modes are unmatched in terms of field enhancement and spatial confinement, and dynamical all-optical control can be achieved e.g. by employing phase-change materials. However, the excitation of surface polaritons in planar structures is intrinsically limited to p-polarization. On the contrary, waveguide modes in high-permittivity films can couple to both p- and s-polarized light, and in thin films, their confinement can become comparable to surface polaritons. Here we demonstrate that the s-polarized waveguide mode in a thin $\text{Ge}_3\text{Sb}_2\text{Te}_6$ (GST) film features a similar dispersion, confinement, and electric field enhancement as the SPhP mode of the silicon carbide (SiC) substrate, while even expanding the allowed frequency range. Moreover, we experimentally show that switching the GST film grants non-volatile control over the SPhP and the waveguide mode dispersions. We provide an analytical model for the description of the PCM/SiC waveguide mode and show that our concept is applicable to the broad variety of polar crystals throughout the infrared spectral range. As such, complementarily to the polarization-limited surface polaritons, the s-polarized PCM waveguide mode constitutes a promising additional building block for nanophotonic applications.
\end{abstract}



\section{Introduction}

The field of nanophotonics pursues control over near-field optical phenomena that can be tailored in custom-designed nanoscale material systems. Key to achieve this control are surface-bound modes such as surface plasmon polaritons (SPP) in metals\cite{Maier2007} or surface phonon polaritons (SPhP) in polar crystals\cite{Caldwell2015}, which stand out by their unprecedented ability to confine and enhance the optical field into subwavelength structures\cite{Kauranen2012}. In the past decades, the potential of these modes has been demonstrated by the large number of applications found in the fields of spectroscopy and sensing\cite{Homola1999,LeRu2012,Berte2018}, photonic circuitry\cite{Engheta2007,DeglInnocenti2018}, lasers\cite{Bergman2003,Berini2012}, and nonlinear optical phenomena\cite{Hillenbrand2002,Kauranen2012,Passler2019}. 

SPPs and SPhPs are electromagnetic evanescent modes bound to the interface of two materials with permittivities $\text{Re}(\varepsilon)$ of opposite sign. SPPs are supported in metals or doped semiconductors from infrared (IR) up to visible frequencies, where the negative $\text{Re}(\varepsilon)$ arises from the Drude response of free electrons, whereas SPhPs arise in the mid to far IR range in the reststrahlen band of polar crystals between the transverse optical (TO) and longitudinal optical (LO) phonon frequencies. In comparison, SPhPs are valued for their longer lifetimes than SPP due to the low losses of the driving phonon resonances\cite{Caldwell2015}. Fundamentally, however, both modes consist of collective charge oscillations coupled to light, leading to shared properties such as subwavelength confinement, strong field enhancement, an asymptotic dispersion, and waveguiding characteristics, which are the reasons for the versatile usability of polaritonic modes. 

However, SPPs and SPhPs in planar heterostructures intrinsically carry out-of-plane electric fields, and thus can only be excited by p-polarized light, while no s-polarized light can couple to them. This polarization restriction limits the versatility of polariton-based nanophotonics and potentially hinders the development of future technologies, such as emission control\cite{Jun2011,Ito2014,Nakamura2016} or exploiting solar energy\cite{Catchpole2008,Ferry2010}. Very recently, omni-polarization waveguide modes in high-permittivity planar media were proposed to circumvent the polarization bottleneck of polaritons\cite{Papadakis2019}, showing that in thin films, waveguide modes can reach comparable degrees of confinement as polaritons. However, while gaining a waveguide mode capable of coupling to s-polarized light, a high-permittivity thin film does not support a SPP or SPhP.

Another drawback of surface polaritons is their lack of active tunability when considered in a plain metal (polar crystal), because there the SPP (SPhP) dispersion is bounded by the fixed plasma frequency (optical phonon frequencies). For SPPs, this disadvantage can be overcome in doped semiconductors\cite{Runnerstrom2019} or graphene\cite{Ni2016,Fei2012}, where the photo- or voltage-induced free-carrier concentration enables tuning of the plasma frequency. A similar approach has been proposed for SPhPs\cite{Dunkelberger2018}, but here, the tunability of the SPhPs gained by an additional photo-induced plasma contribution is limited. 

A significantly larger non-versatile switching contrast can be achieved by utilizing phase-change materials (PCMs)\cite{Wuttig2008,Wuttig2018,Raty2019} such as the $\text{Ge}_3\text{Sb}_2\text{Te}_6$ compound (GST-326), which can be reversibly switched between its amorphous (a-) and crystalline (c-) phases by ultrafast electrical or optical pulses\cite{Wuttig2007,Bruns2009,Gholipour2013,Waldecker2015,Wang2016,Michel2019}. The high permittivity contrast between the two phases ($\varepsilon_{\text{a-GST}} \approx 13.8 + 0i$ and $\varepsilon_{\text{c-GST}} \approx 31.4 + 9i$; values are obtained from a global fitting procedure, see Methods section for details) in the IR is sensed by SPhPs that evanescently penetrate the GST, leading to a significant shift of the SPhP mode frequency upon switching of the GST phase\cite{Li2016}. However, while systems utilizing PCMs have enabled active control over surface polaritons, the large potential of s-polarized waveguide modes in these systems remains, so far, unexploited.

\section{Concept}

In this work, we suggest a generic, actively tunable material system for the infrared spectral range that simultaneously supports p-polarized SPhPs and s-polarized waveguide modes. Intriguingly, the waveguide modes replicate the properties of the SPhP that define its suitability for nanophotonic applications, such as the enhancement of local electric fields and subwavelength confinement. We experimentally demonstrate active tuning of both excitations by means of a phase-change material, which, at the same time, constitutes the waveguide core. As a model system, we analyze the tuning of SPhP and waveguide modes in a GST/SiC structure, where the GST film can be switched between its amourphous and crystalline state. Offering broad functionality, such as polarization dependent polariton optics\cite{Folland2018,Chaudhary2019}, the proposed material system features a unique combination of active tunability of its guided modes and the lifting of the polarization bottleneck of conventional polaritonics.

\begin{figure}
\includegraphics[width=.9\linewidth]{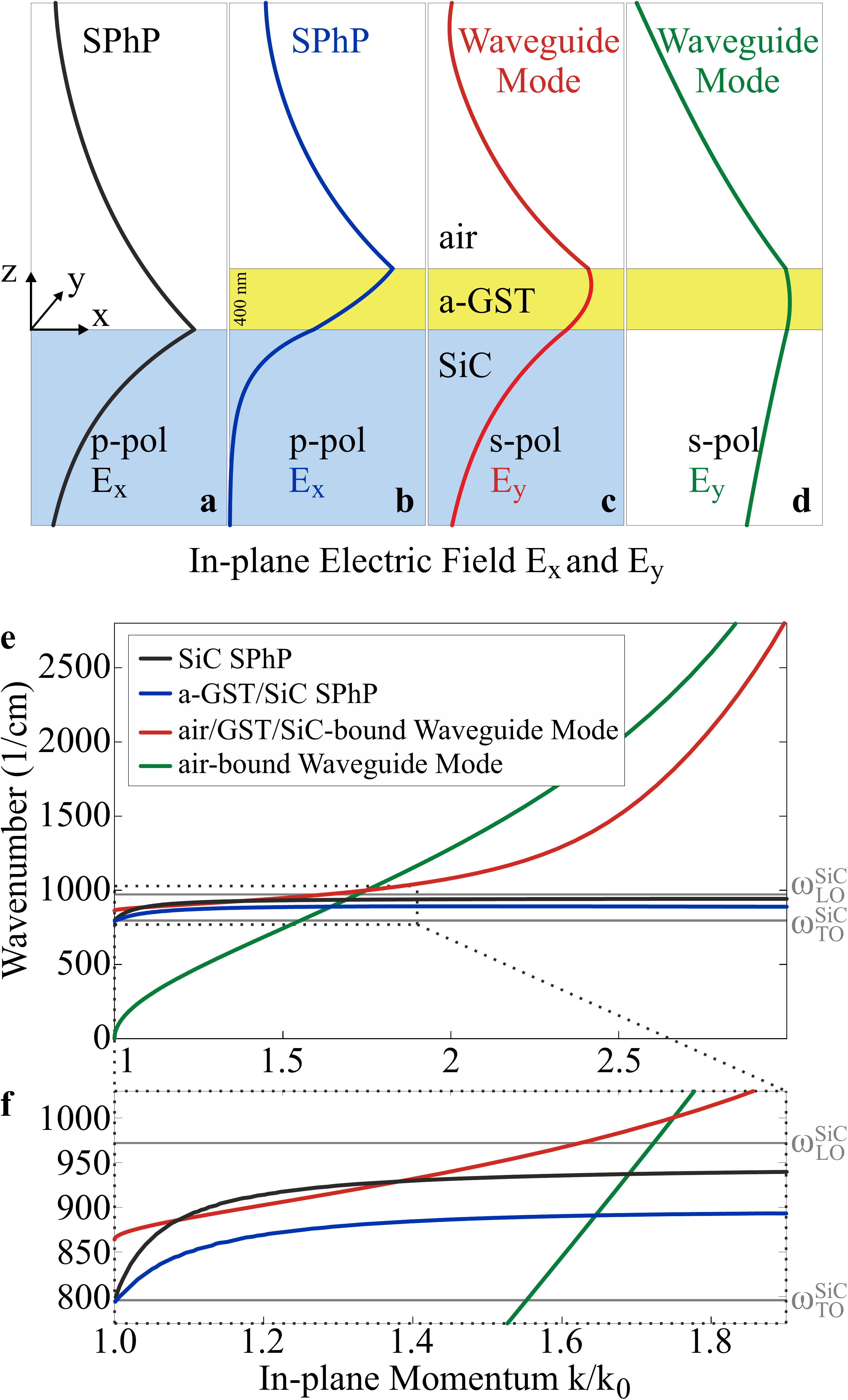}
  \caption{\textbf{Guided modes in an a-GST film on a SiC substrate.} \textbf{a-d} Schematic in-plane electric field distributions of SPhPs and waveguide modes, excitable with p- and s-polarized light, respectively, supported on a bare SiC substrate (a), in a \nmetr{400} thin a-GST film on SiC (b-c), and in a freestanding a-GST film (d). \textbf{e-f} Dispersions of the four modes shown in a-d, where f zooms into the SiC reststrahlen region between the TO and LO frequencies ($\omega_{TO}$ and $\omega_{LO}$), marked by the dotted rectangle. The SiC substrate leads to a flattening of the waveguide mode (red line) compared to the air-bound mode (green line), leading to a comparable dispersion to the SPhP (black and blue lines) in the dispersion window shown in f.}
  \label{fig1}
\end{figure}

A SPhP propagating along the surface of a polar crystal such as SiC exhibits an in-plane $E_x$ field that peaks at the interface, see Fig. \ref{fig1}a. We define the $x$-$z$ plane as the plane of incidence, and all waves propagate in $x$-direction. In Fig. \ref{fig1}e (and zoomed into the reststrahlen region of SiC in Fig. \ref{fig1}f), the dispersion of this SPhP is shown in black. By placing a \nmetr{400} thin a-GST onto SiC, the SPhP dispersion is red-shifted (blue line in Fig. \ref{fig1}e,f), while maintaining the shape of its in-plane electric field $E_x$, see Fig. \ref{fig1}b. Interestingly, here, the maximum field is encountered at the air/a-GST interface, and decays exponentially through the GST film into the SiC substrate (please note that for illustration purposes, the GST film thickness in Fig. \ref{fig1}a-d is not to scale with respect to the shown z-range of the adjacent air and the substrate).

A freestanding \nmetr{400} thin a-GST film, on the other hand, supports an s-polarized waveguide mode owing to the high index contrast between a-GST and air\cite{Tien1970}. The in-plane electric field $E_y$ of this mode is shown in Fig. \ref{fig1}d, and the corresponding dispersion is plotted in green in Fig. \ref{fig1}e,f. Notably, this mode features a much larger penetration depth $\delta$ in positive and negative z-direction, with $E(z)=E(0) e^{-z/\delta}$, than the SPhP, hence it is much less confined within the a-GST film. The dispersion of this waveguide mode, however, is steeper and covers a much larger frequency range than the SPhP, which is restricted to the SiC reststrahlen band. Now, the combined system of a-GST/SiC supports not only the shifted p-polarized SPhP mode but simultaneously an s-polarized waveguide mode localized in the a-GST film. In Fig. \ref{fig1}c, the in-plane electric field $E_y$ of this air/GST/SiC-bound mode is shown, featuring a higher confinement than the air-bound waveguide mode due to the SiC substrate. Strikingly, the dispersion of the air/GST/SiC-bound waveguide mode (red line in Fig. \ref{fig1}e,f), while approaching asymptotically the dispersion of the air-bound waveguide mode at large in-plane momenta $k$, experiences an upward bending for small $k$, becoming quite similar to the dispersion of the SPhP. Thus, the combined structure of a thin a-GST film on a SiC substrate not only supports both p-polarized SPhPs and s-polarized waveguide modes, but thanks to the reststrahlen characteristics\cite{Paarmann2015} of the SiC substrate, the waveguide mode mimicks the SPhP for small in-plane momenta $k/k_0$ in both field distribution and dispersion.

\section{Experiment}

\begin{figure}
\includegraphics[width=.9\linewidth]{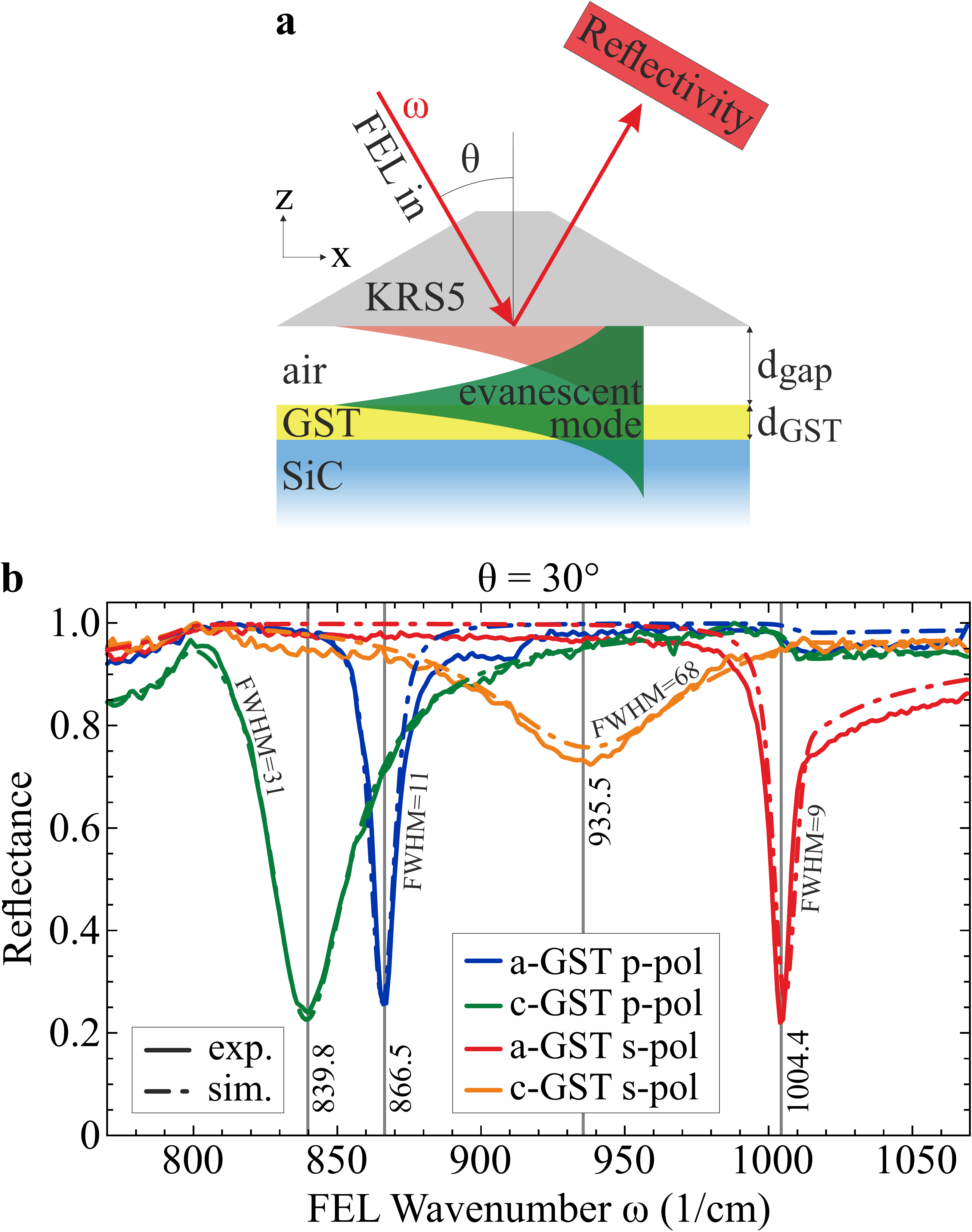}
  \caption{\textbf{Tuning of SPhP and waveguide mode by switching the GST phase.} \textbf{a} Setup of the prism coupling experiments in the Otto geometry with variable air gap between the KRS5 ($n_{KRS5}\approx 2.4$) prism and the sample, allowing for momentum-matched, critical coupling to evanescent modes supported by the sample such as SPhPs and waveguide modes. \textbf{b} Reflectance spectra at $\theta=\dg{30}$ for the crystalline and amorphous GST phases for s- and p-polarized incident light of a \nmetr{144} (\nmetr{132}) thin a-GST (c-GST) film on SiC, demonstrating the tuning of both SPhP and waveguide mode upon switching the GST phase. A global fitting algorithm was used to fit simulated reflectances (dash-dotted lines) to the experimental spectra (solid lines).}
  \label{fig2}
\end{figure}

\begin{figure*}[t]
\includegraphics[width=\linewidth]{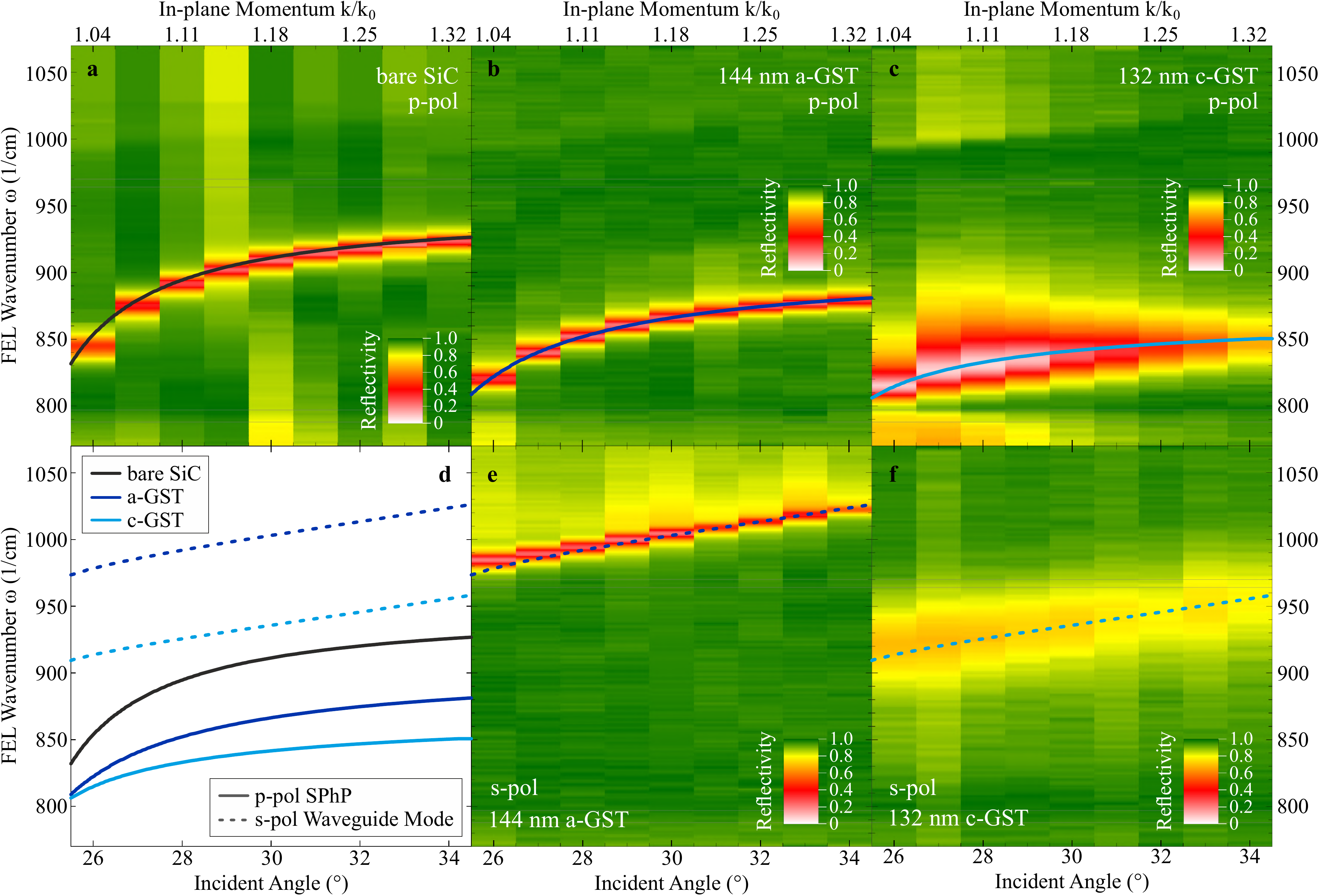}
  \caption{\textbf{Experimental reflectance maps featuring the SPhP and waveguide mode dispersions.} \textbf{a-c} Reflectance maps for p-polarized incident light of bare SiC, a-GST/SiC, and c-GST/SiC, respectively, demonstrating the red-shift of the SPhP dispersion induced by the sensing of the increasing refractive index from air to c-GST. \textbf{e-f} Reflectance maps for s-polarized incident light, showing the red-shift of the waveguide mode supported by the \nmetr{144} (\nmetr{132}) thin a-GST (c-GST) film. \textbf{d} Summary of all mode dispersions determined by the reflectance measurements, also plotted on top of the corresponding reflectance map.}
  \label{fig3}
\end{figure*}

In a GST film of $\sim\nmetr{400}$ thickness, as discussed in Fig. \ref{fig1}, the waveguide mode features maximal field enhancement. However, a maximum frequency shift upon switching of the GST phase is obtained for a GST film thickness of $\sim\nmetr{150}$, while maintaining a significant field enhancement (discussed in detail in Fig. \ref{fig5}). Thus, to demonstrate the concept, in our work a \nmetr{144} thin a-GST film was sputter-deposited onto a 4H-SiC substrate and placed into a home-build Otto-type prism coupling setup, sketched in Fig. \ref{fig2}a (see Methods). In short, the incoming light impinges on the prism surface at an angle $\theta$ above the critical angle of total internal reflection ($\sim \dg{24.6}$ for KRS5). The generated evanescent wave in the air gap between prism and sample couples to the evanescent, bound modes in the sample, resulting in attenuated total reflection dips at the resonances in a reflectance spectrum. By taking several spectra at varying incident angle, dispersions can be mapped out\cite{Passler2017}. As an excitation source for the reflectance spectroscopy measurements, we employed a mid-IR free electron laser (FEL) with tunable wavelength in the range of $3-\mumetr{60}$ and a spectral line width of $\sim 0.3\%$\cite{Schollkopf2015} (see Methods). In a first step, all experimental data were obtained for a-GST, and after crystallizing the a-GST on a heating plate ($\dg{300}\text{C}$, 5 min), the measurements were repeated for the c-GST phase. Note that crystallization of a-GST is accompanied by a density reduction\cite{Njoroge2002,Michel2016a}, leading to a $\sim 8\%$ thinner c-GST film of about \nmetr{132} thickness.

In Fig. \ref{fig2}b, reflectance spectra at an incident angle of \dg{30} are shown for p-polarized and s-polarized incident light and for both GST phases, respectively. Clearly, each spectrum features a distinct resonance dip that corresponds to a p-polarized SPhP (blue and green curves) or an s-polarized waveguide mode (red and orange curves). All data were fitted with a global fitting procedure (see Methods for details), and the resulting theoretical spectra calculated by means of a $4\times 4$ transfer matrix formalism\cite{Passler2017a} are also shown in Fig. \ref{fig2}b, nicely reproducing the experimental data. (The low amplitude of the waveguide mode in c-GST arises due to a prism coupling gap below the gap of critical coupling\cite{Passler2017}, see Methods for details.)

Furthermore, a Lorentzian line shape was fitted to the experimental reflectance dips in order to extract frequency positions $\mu$ and full width at half maxima ($\text{FWHM}=2 \sigma$) of the resonances. Interestingly, the measured quality factor ($Q=\mu/\text{FWHM}$) in a-GST of the waveguide mode ($Q=109$) is even higher than that of the SPhP ($Q=82$), whereas in c-GST the waveguide mode has a broader resonance ($Q=14$) than the SPhP ($Q=27$). For the case of a-GST, these Q-factors easily lie in the range of the high-Q resonances arising from localized SPhPs in polar crystal nanostructures (45-150)\cite{Wang2013,Caldwell2013,Sumikura2019}. The generally lower $Q$-factors in c-GST, on the other hand, can be attributed to non-zero absorption ($\text{Im}\left(\varepsilon_{\text{c-GST}}\right)=9$) compared to a-GST. However, it is noteworthy that the waveguide mode features similar $Q$-factors as the SPhP in both GST phases, providing first evidence that the waveguide mode can indeed be seen as a polariton-like excitation with s-polarization.

Reflectance spectra were taken for incident angles of $26-\dg{34}$ in steps of \dg{1} resulting in reflectance maps that reproduce the dispersion of the corresponding mode. Maps for bare SiC, a-GST/SiC and c-GST/SiC for p-polarized incident light are shown in Fig. \ref{fig3} a-c, respectively, and maps for a-GST/SiC and c-GST/SiC for s-polarized incident light in Fig. \ref{fig3}e-f. In Fig. \ref{fig3}d, an overview over all modes is given, and the respective dispersion curves are additionally plotted on top of the individual reflectance maps. The theoretical dispersions were extracted from the imaginary part of the correspondingly polarized reflection coefficient obtained by transfer matrix calculations\cite{Passler2017a} using the parameters from the global fitting procedure. All experimental dispersions are well reproduced by the simulations. 

For a \nmetr{144} thin a-GST and at $\theta=\dg{30}$, the SPhP shifts by $\delta\omega=\wavenumber{27}$ upon switching the GST phase, and the waveguide mode by \wavenumber{69}. This corresponds to a tuning figure of merit (TFOM) of $\text{TFOM}=\delta\omega/\text{FWHM}=2.5$ for the SPhP and 7.7 for the waveguide mode, where we used the FWHM of the a-GST phase, yielding a better TFOM than for the c-GST phase. These TFOM by far exceed other recently published results in the field of modulated nanophotonics with typical values of 0.5-1.4\cite{Yao2013,Si2015,Dunkelberger2018,Michel2019}, and, in the case of the waveguide mode with an exceptional TFOM of 7.7, even exceed the results on graphene, where values up to 5 have been reported\cite{Fang2013,Gao2013}. In the experimental reflectance maps in Fig. \ref{fig3}, this switching can be observed as a drastic red-shift of the corresponding dispersion curves. The GST film thickness was chosen such that the frequency shift is maximal for both the SPhP and the waveguide mode. Indeed, for thicker films, the shift decreases, because all modes for both GST phases are pushed against the lower frequency limit given by $\omega_{TO}^{\text{SiC}}=\wavenumber{797}$\cite{Engelbrecht1993}. For thinner films, on the other hand, the SPhP contrast decreases as well because of a reduced influence of the GST film, while the waveguide mode disappears completely. The film thickness dependence and the boundaries in which the waveguide mode is supported are discussed further below, see Fig. \ref{fig4}. For the exemplary film thickness of $\nmetr{144}$, however, our experimental findings clearly demonstrate the tuning potential of both the p-polarized SPhP and the s-polarized waveguide mode supported in GST films on a SiC substrate via GST phase switching.

\section{Waveguide Mode Theory}

In the following, we provide a theoretical description of the s-polarized waveguide mode following Tien and Ulrich\cite{Tien1970} in order to verify its waveguide mode nature, as well as provide a thorough understanding of the dispersion of this mode in the GST/SiC structure and the boundaries in which it is supported. We will mainly focus our discussion on a-GST, which is the GST phase with lower losses, thus featuring better waveguide characteristics. We note, however, that the following theoretical description is equally valid for c-GST. In fact, except for a red-shift of the dispersion and a larger momentum range, the resulting dispersion for c-GST is qualitatively the same as for a-GST (except at the upper momentum boundary, see Supporting Information Figure S1 for details).

Solutions of confined, propagating electromagnetic waves in a thin film, i.e. waveguide modes, can be found by searching for plane waves that interfere constructively within the film boundaries. For the mode to be confined to the film, the refractive index $n_{\text{GST}}$ of the film has to be larger than those of the adjacent substrate $n_{\text{SiC}}$ and the incident medium $n_{\text{air}}$ ($n_{\text{GST}}>n_{\text{SiC}},n_{\text{air}}$), such that the mode can propagate in the film but is evanescent in both adjacent media. The resulting wave performs a zig-zag motion in the thin film and experiences total internal reflection at both film interfaces. The phase differences $\Phi_{\text{SiC/GST}}$ and $\Phi_{\text{GST/air}}$ upon reflection at the substrate/film (SiC/GST) and the film/incident medium (GST/air) interfaces depend on the in-plane momentum $k/k_0=n_{\text{air}} \, \text{sin} \:\theta \equiv \kappa$ (with $k_0=\omega/c_0$) and on the corresponding material permittivities, and are given by\cite{Tien1970}:
\begin{align}
\text{tan}\left( \Phi_{\text{SiC/GST}}^s \right) &= \sqrt{\frac{\kappa^2-\varepsilon_{\text{SiC}}}{\varepsilon_{\text{GST}}-\kappa^2}}, \label{eq:phi} \\
\quad \text{tan}\left( \Phi_{\text{GST/air}}^s \right) &= \sqrt{\frac{\kappa^2-\varepsilon_{\text{air}}}{\varepsilon_{\text{GST}}-\kappa^2}}, \\
\text{tan}\left( \Phi_{\text{SiC/GST}}^p \right) &= \frac{\varepsilon_{\text{GST}}}{\varepsilon_{\text{SiC}}} \sqrt{\frac{\kappa^2-\varepsilon_{\text{SiC}}}{\varepsilon_{\text{GST}}-\kappa^2}}, \\ 
\text{tan}\left( \Phi_{\text{GST/air}}^p \right) &=\frac{\varepsilon_{\text{GST}}}{\varepsilon_{\text{air}}} \sqrt{\frac{\kappa^2-\varepsilon_{\text{air}}}{\varepsilon_{\text{GST}}-\kappa^2}},
\end{align} 

where the superscripts $s$ and $p$ indicate the polarization, $0 \! \leq \! \Phi \! \leq \! \pi/2$, and, in order to ensure total internal reflection, the phase differences $\Phi$ have to be real. A solution of a waveguide mode is found when the phase difference of a plane wave after travelling one zig-zag path is $2 m\pi$, with $m$ being the order of the waveguide mode. The equation of modes then is\cite{Tien1970}:
\begin{align}
\frac{\omega}{c_0} \sqrt{\varepsilon_{\text{GST}}-\kappa^2} \; d - \Phi_{\text{SiC/GST}} - \Phi_{\text{GST/air}} = m \pi,
\label{eq:modes}
\end{align}

where $d$ is the film thickness, $\omega$ is the frequency, and $c_0$ the speed of light in vacuum. Please note that Eq. \ref{eq:modes} is valid for s- and p-polarized waveguide modes. In this work, however, we focus on the zero-order s-polarized waveguide mode and its comparibility to the simultaneously supported SPhP mode. The following discussion will therefore only consider the s-polarized waveguide solutions.

\begin{figure*}[t]
\includegraphics[width=\linewidth]{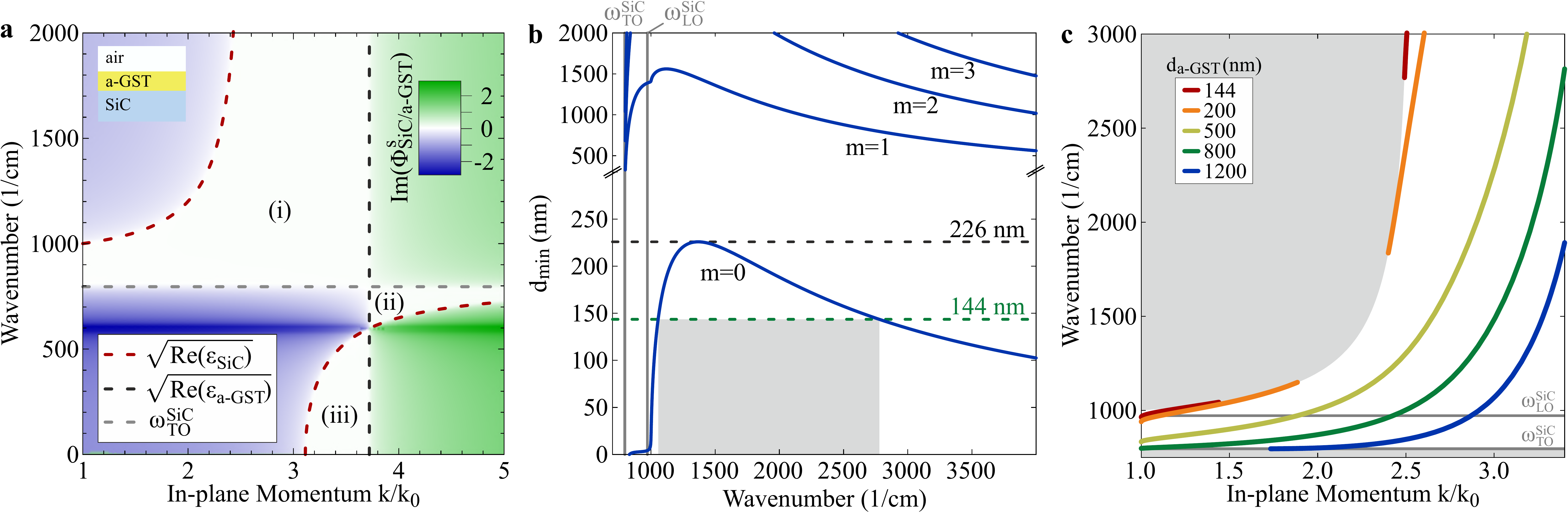}
  \caption{\textbf{Theoretical dispersion of an s-polarized waveguide mode in an air/a-GST/SiC system.} \textbf{a} Imaginary part of the phase difference $\Phi_{\text{SiC/GST}}^s$ (Eq. \ref{eq:phi}) upon reflection of s-polarized light at the a-GST/SiC interface. White areas where $\text{Im}(\Phi_{\text{SiC/GST}}^s)=0$ indicate regions where the a-GST/SiC interface supports a waveguide mode. Dashed lines mark the respective boundaries given by material parameters. \textbf{b} Minimum a-GST film thickness (Eq. \ref{eq:dmin}) required to support an m-th order s-polarized waveguide mode. The $m=0$ mode is supported at all frequencies for film thicknesses above \nmetr{226}. For the a-GST film of \nmetr{144} thickness employed in our experiments, the mode cannot exist in the frequency window of $1056-\wavenumber{2769}$ marked in gray. \textbf{c} Theoretical dispersion (Eq. \ref{eq:modes}) of the s-polarized $m=0$ air/GST/SiC-bound waveguide mode. The gray areas are the forbidden regions for the air/a-GST/SiC system given by $\text{Im}( \Phi_{\text{SiC/GST}}^s )$. The two thinnest films have forbidden frequency regions given by $d_{\text{min}}$ (b), coinciding with the restriction given by $\text{Im} ( \Phi_{\text{SiC/GST}}^s )$.}
  \label{fig4}
\end{figure*}

As a consequence of the requirement of the phase differences $\Phi$ to be real, the imaginary part of $\Phi$ allows to define those regions in the dispersion where a waveguide mode is supported, and those in which mode propagation is not possible. Fig. \ref{fig4}a shows a map of $\Phi_{\text{SiC/GST}}^s$ for the air/a-GST/SiC system. The three white areas (i)-(iii) are the regions where the a-GST/SiC interface supports the s-polarized waveguide mode ($\text{Im}(\Phi_{\text{SiC/a-GST}}^s)=0$). However, for a mode to be supported in the film, the phase differences at both interfaces have to be real simultaneously. Therefore, region (ii) can be ruled out because for in-plane momenta $\kappa  \! > \! \sqrt{\text{Re}(\varepsilon_{\text{a-GST}})} \! \approx \! 3.7$ (black dashed line), the phase difference at the a-GST/air interface is complex, $\text{Im}(\Phi_{\text{a-GST/air}}^s) >0$ (see Supporting Information Figure S1). For $\kappa< 3.7$, on the other hand, $\text{Im}(\Phi_{\text{a-GST/air}}^s)=0$ and thus regions (i) and (iii) can support waveguide modes, and their boundaries are solely governed by the a-GST/SiC interface. In fact, the lower boundary along the momentum axis is given by the SiC permittivity $\sqrt{\text{Re}(\varepsilon_{\text{SiC}})}$ (red dashed line), which is a highly dispersive quantity and is the reason for the rich physics of waveguide modes bounded by polar crystal substrates. While in principle region (iii) does support waveguide modes, we here only focus on region (i) due to its capacity of supporting the s-polarized waveguide mode that features an upwards bended, polariton-like dispersion in proximity to the SiC substrate SPhP mode (see Fig. \ref{fig1}e,f). As for the SPhP, the lower boundary of region (i) along the frequency axis is defined by the TO phonon frequency of SiC, $\omega_{TO}^{\text{SiC}}$ (gray dashed line). Interestingly, the reason that $\omega_{TO}^{\text{SiC}}$ is a boundary is different for the two modes: The SPhP requires negative $\text{Re}(\varepsilon_{\text{SiC}})$ to have an evanescently decaying electric field in z-direction. On the other hand, the waveguide mode requires $\text{Re}(\varepsilon_{\text{SiC}}) <\text{Re}(\varepsilon_{\text{a-GST}})$ to enable total internal reflection and thereby get evanescently decaying fields. For frequencies $\omega<\omega_{TO}^{\text{SiC}}$, both conditions are lost simultaneously.

Additional to the restrictions given by total internal reflection that are reflected in the phase differences $\Phi$ (Fig. \ref{fig4}a), waveguide modes require a minimal film thickness $d_{\text{min}}$ in order to be supported. Evaluation of Eq. \ref{eq:modes} at the lower and upper momentum boundaries leads to the following expression for $d_\text{{min}}$ for s-polarized waveguide modes\cite{Tien1970}:
\begin{align}
d_{\text{min}}=\frac{c_0}{\omega} \: \frac{m\pi+ \text{arctan} \left( \sqrt{\frac{\varepsilon_{\text{SiC}}-\varepsilon_{\text{air}}}{\varepsilon_{\text{GST}} - \varepsilon_{\text{SiC}}}} \right)}{\varepsilon_{\text{GST}} - \varepsilon_{\text{SiC}}}.
\label{eq:dmin}
\end{align}


In Fig. \ref{fig4}b, $d_{\text{min}}$ of the air/a-GST/SiC system is plotted as a function of frequency for the first four orders $m$. Interestingly, in the SiC reststrahlen band, $d_{\text{min}}$ of the $m\!=\!0$ mode is almost zero, but features a rapid increase at larger frequencies with a maximum value of $d_{\text{min}}=\nmetr{226}$. As a consequence, a-GST films of thicknesses $d_{\text{a-GST}}>\nmetr{226}$ support the $m\!=\!0$ mode at any frequencies, while in thinner films, such as our sample with $d_{\text{a-GST}}=\nmetr{144}$, the mode is forbidden in a broad frequency range (gray area in Fig. \ref{fig4}b). Therefore, the $m \! =\! 0$ s-polarized waveguide mode we observe in our experiments is limited to a frequency range between $797-\wavenumber{1056}$, which is very similar to the allowed frequency range of the SPhP defined by the SiC reststrahlen band ($797-\wavenumber{970}$). 

The range in which the s-polarized air/a-GST/SiC waveguide mode can be observed is best illustrated in the dispersion plot we show in Fig. \ref{fig4}c. Here, the gray areas indicate the forbidden regions given by $\text{Im}(\Phi_{\text{SiC/a-GST}}^s)$. For the two thinnest films of $144$ and \nmetr{200} (red and orange curves), $d_{\text{min}}$ defines a frequency range where the $m\!=\!0$ mode is forbidden, which is represented by the region in which the two dispersion curves are disrupted. The restriction given by $d_{\text{min}}$ coincides exactly with the restriction given by $\text{Im}(\Phi_{\text{SiC/a-GST}}^s)$. For the three thicker films (light green, dark green, and blue curves), on the other hand, the mode dispersions are continuous, as there is no minimum film thickness required. However, these mode dispersions are pushed against the lower frequency boundary $\omega_{TO}^{\text{SiC}}$ with increasing film thickness, and the dispersion of the mode in the thickest film of \nmetr{1200} even starts to disappear at this boundary. This behavior originates in the pole of the dielectric function of SiC at $\omega_{TO}^{\text{SiC}}$, constituting the hard lower barrier of the frequency range in which waveguide modes can be supported. Notably, thanks to this boundary, the waveguide mode dispersion experiences the upward bending that leads to the resemblence of the SPhP mode dispersion. Finally, we note that the shifting of the dispersion with film thickness enables a straightforward way of tuning the mode resonance across the entire allowed frequency range. 




\begin{figure*}[t!]
\includegraphics[width=.99\linewidth]{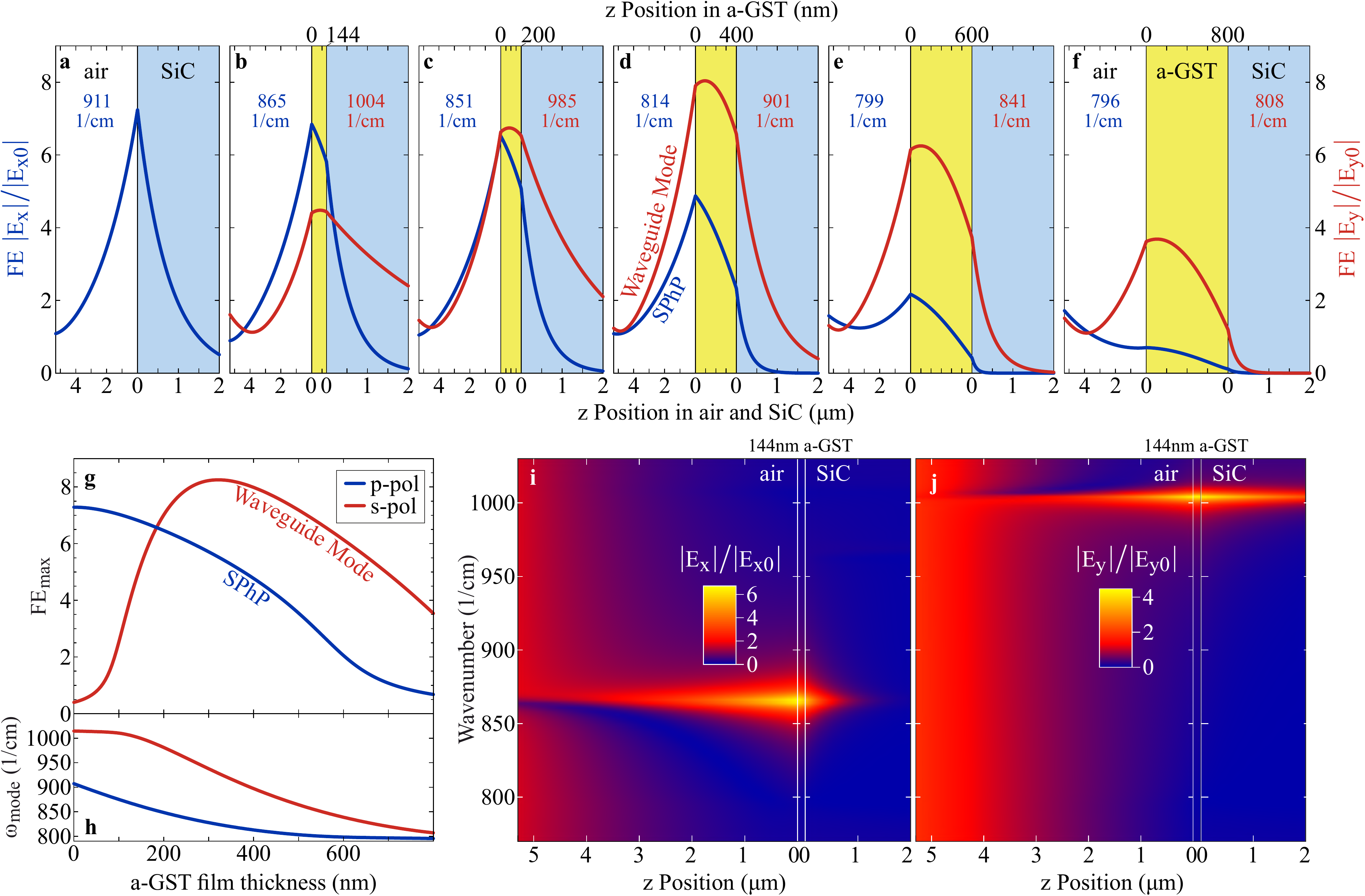}
  \caption{\textbf{Electric field enhancement of a SPhP and a waveguide mode in an a-GST film.} \textbf{a-f} In-plane electric field profiles along the z-axis perpendicular to the air/a-GST/SiC interfaces of the SPhP (blue lines, $E_x$ due to p-polarization) and the waveguide mode (red lines, $E_y$ due to s-polarization) for a series of a-GST film thicknesses. The field enhancements (FEs) are calculated within the Otto geometry for an in-plane momentum of $k/k_0=1.18$ and at the respective resonance frequency (specified in each graph). Note the different scales of the z-axis for air and SiC (bottom scale) and the GST film (top scale), which are used to cover all relevant length scales. \textbf{g,h} Maximum field enhancement FE$_{\text{max}}$ inside the a-GST film (g) as a function of a-GST film thickness for s- (red curve) and p-polarization (blue curve) calculated at the respective mode frequencies (h). \textbf{i,j} Spatio-spectral maps of the in-plane electric field enhancements of the p-polarized SPhP and the s-polarized waveguide mode, respectively, showing their different resonance frequencies while featuring comparable Q-factors. The maps are also calculated for an in-plane momentum of $k/k_0=1.18$. In contrast to a-f, we here plot with a single scale of the z-axis in order to better visualize the evanescent depth of the modes along the z-direction. The z-positions are labelled relative to the respective interface with a-GST.}
  \label{fig5}
\end{figure*}

\subsection{Surface Polariton-like Waveguide Modes for s-Polarized Nanophotonics}

Equipped with this theoretical understanding, we now turn our focus on the comparability of the SPhP and the thin-film s-polarized waveguide mode in terms of optical field enhancement and spatial confinement. Fig. \ref{fig5} a-f show z-profiles of the in-plane electric field enhancement (FE) of the SPhP ($E_x$) and the waveguide mode ($E_y$) at their corresponding resonance frequency for a set of different a-GST film thicknesses, calculated for excitation via a KRS5 coupling prism in the Otto geometry at an incident angle of $\theta=\dg{30}$. Furthermore, the continuous behavior of the of the maximum field enhancement FE$_{\text{max}}$ for the SPhP (blue curve) as a function of thin a-GST film thickness $d_{\text{a-GST}}$ is shown in Fig. \ref{fig5}g, and Fig. \ref{fig5}h shows the corresponding mode frequency $\omega_{\text{mode}}$ (FE$_{\text{max}}$ and $\omega_{\text{mode}}$ for larger film thicknesses are shown in Supporting Information Figure S2). 

A bare SiC surface (Fig. \ref{fig5}a) only supports a SPhP with a maximum field enhancement (FE$_{\text{max}}$) of $\sim7$, which is, however, reduced when an additional a-GST film of \nmetr{144} is placed on top of SiC (Fig. \ref{fig5}b). For increasing film thicknesses (Fig. \ref{fig5}c-e), the field enhancement of the SPhP drops even further and for the largest a-GST film of \nmetr{800} thickness (Fig. \ref{fig5}f) there is no enhancement observable at all, meaning that the structure does not support the air/SiC SPhP anymore. This transition towards a total loss of the SPhP for large film thicknesses stems, on the one hand, from the fact that the observed air/SiC SPhP originates in the dielectric contrast of the adjacent air and SiC, and for growing a-GST film thicknesses, there is no SiC/air interface anymore. On the other hand, the TO phonon is IR-active and therefore absorption in SiC increases drastically in the vicinity of $\omega_{TO}^{\text{SiC}}$, contributing to the observed loss of field enhancement. Finally, we cannot observe the emergence of any other SPhP for thicker films, because the air/a-GST interface does not feature permittivities of opposite sign, while a potential a-GST/SiC SPhP cannot be excited via prism coupling with a KRS5 prism, because the required momenta are too large ($n_{\text{KRS5}}<n_{\text{a-GST}}$). 

The s-polarized waveguide mode, on the other hand, features a maximum field enhancement of $\sim 8$ in a \nmetr{300} thin a-GST film, which decreases for both thinner and thicker films, see Fig. \ref{fig5}g (red curve). In order to explain this behavior, we recall that for small in-plane momenta ($1\!<\!\kappa\!<\! 2$), the waveguide mode is restricted to a frequency range between $\omega_{TO}^{\text{SiC}}=\wavenumber{797}$ and $1000-\wavenumber{1200}$, where the upper boundary depends on $\kappa$, see Fig. \ref{fig4}c. With increasing film thickness, the dispersion red-shifts through the allowed frequency region and eventually is pushed against $\omega_{TO}^{\text{SiC}}$, as can be observed in Fig. \ref{fig5}h (red curve). Here, analogeous to the SPhP, the high absorption in SiC close to $\omega_{TO}^{\text{SiC}}$ leads to the observed loss of field enhancement. At the upper frequency boundary, on the other hand, the permittivity of SiC is close to 1, resulting in a less confined mode in SiC compared to modes at frequencies further into the reststrahlen band (compare the penetration depth into SiC of the waveguide mode in Fig. \ref{fig5} b and d), thus leading to smaller field enhancements. 

The penetration depth into the SiC substrate $\delta$, and thus the spatial confinement of the modes, is solely a function of the SiC permittivity tensor. Close to $\omega_{TO}^{\text{SiC}}$, both the SPhP and the waveguide mode are strongly confined (with a penetration depth of $\delta \approx \nmetr{100}$ at \wavenumber{800}), whereas at higher frequencies, the confinement decreases and the modes leak further into the SiC substrate ($\delta \approx \mumetr{1.5}$ at \wavenumber{950}, see Supporting Information Figure S2 for details). This frequency dependence of the spatial confinement can also be seen in Fig. \ref{fig5}i,j, where we show spatio-spectral maps of the SPhP and the waveguide mode, respectively, for a \nmetr{144} thin a-GST film, resembling the experimentally studied sample. Clearly, the waveguide mode leaks further into SiC, whereas the SPhP is stronger confined because the frequency of the latter is much closer to $\omega_{TO}^{\text{SiC}}$ than that of the former. The opposite behavior can be observed when the waveguide mode frequency is close to  $\omega_{TO}^{\text{SiC}}$, see Fig. \ref{fig5}f, and the SPhP mode sits at higher frequencies (for example Fig. \ref{fig5}a,b).

\begin{figure}
\includegraphics[width=\linewidth]{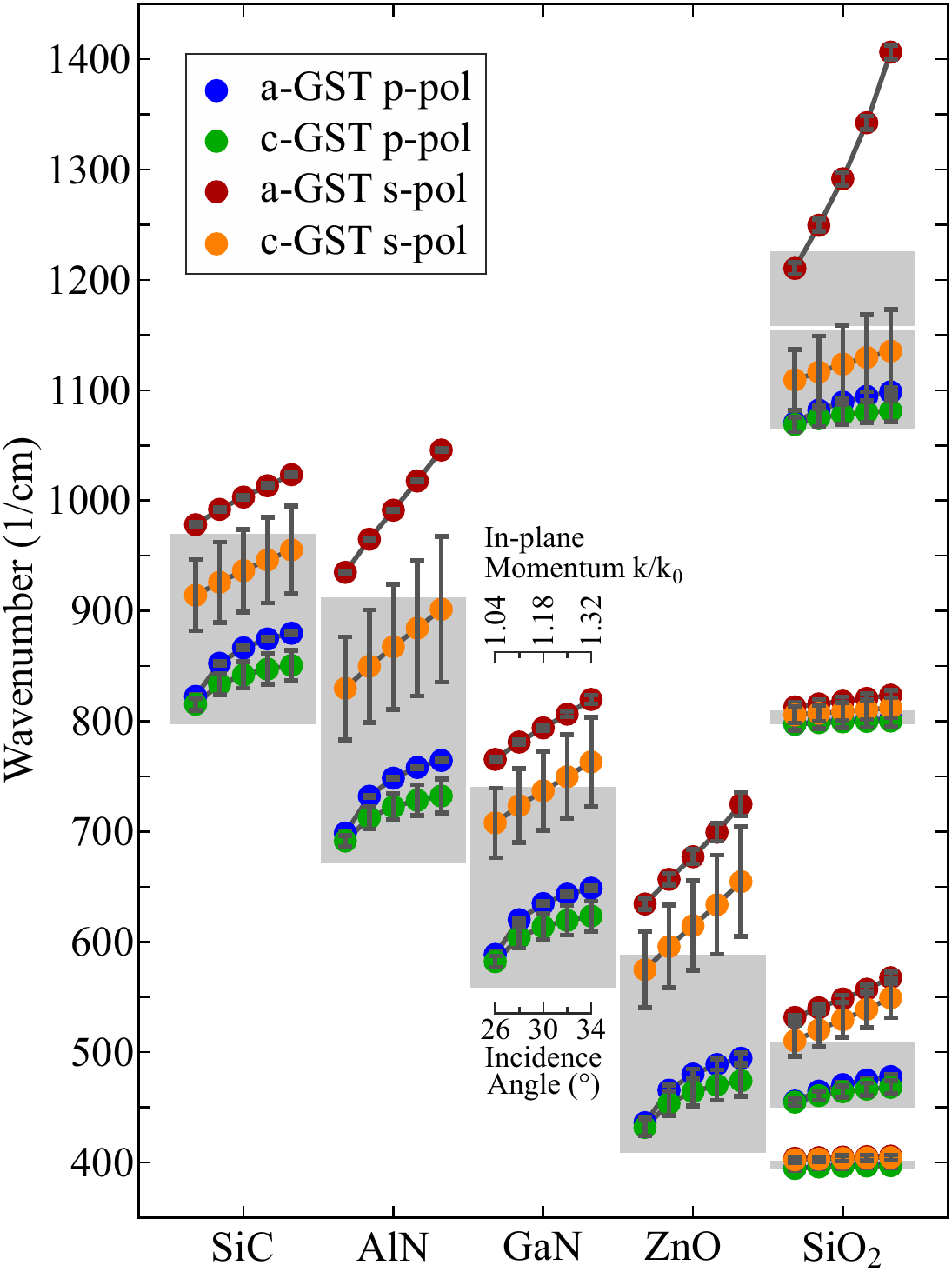}
  \caption{\textbf{Comparison of GST on different polar crystal substrates for supporting s- and p-polarized modes.} SiC, AlN, GaN, and ZnO have a single reststrahlen band (gray areas) and thus support one SPhP (blue, green) and one waveguide mode (red, orange) for each GST phase. The multimode material $\alpha$-SiO$_2$, on the other hand, features four reststrahlen bands and thus four different sets of guided modes. The a-GST (c-GST) film thickness is that of the experimentally studied sample of \nmetr{144} (\nmetr{132}). Resonance positions were extracted from maxima in the imaginary part of the correspondingly polarized reflection coefficient $\text{Im}\left( r^{p,s} \right)$ at the in-plane momenta shown for GaN. The peaks were fitted with a Lorentzian function. The obtained FWHMs are plotted as vertical gray bars on top of each resonance point.}
  \label{fig6}
\end{figure}

As we have shown exemplarily for a SiC substrate, a thin overlying GST film enables an s-polarized waveguide mode with features such as its dispersion, field enhancement, and spatial confinement being comparable to those of the simultaneaously supported SPhP. Strikingly, this concept can be easily transferred to any polar crystal substrate. In Fig. \ref{fig6}, we compare SiC to aluminum nitride (AlN), gallium nitride (GaN), and zinc oxide (ZnO), all featuring a single reststrahlen band, and the multimode material $\alpha$-quartz ($\alpha$-SiO$_2$). For each material, the dispersions of the supported SPhPs (blue and green circles) and s-polarized waveguide modes (red and orange circles) are plotted for a \nmetr{144} thin a-GST and a \nmetr{132} c-GST film, respectively. The vertical gray bars correspond to the FWHM of the mode resonances extracted from calculated spectra of the imaginary part of the corresponding reflection coefficient, $\text{Im}\left( r^{p,s} \right)$. The dispersions are visibly shifted by more than the modes' FWHM between the two GST phases (blue-green and red-orange), clearly demonstrating the good switching contrast achievable in any of the five substrate materials. The gray areas mark the respective restststrahlen bands of SiC, AlN, GaN, and ZnO, and the four bands of $\alpha$-SiO$_2$. In each reststrahlen band, a SPhP within the reststrahlen band and a waveguide mode dispersing above or at the upper edge of the reststrahlen band are supported, highlighting the generality of our concept for the broad IR frequency range of $400-\wavenumber{1400}$. 

\section{Conclusion}
In conclusion, we have presented a generic material system comprising a subwavelength thin PCM film on a polar crystal substrate that simultaneously supports a p-polarized SPhP and an s-polarized waveguide mode. Our experimental data of the exemplary GST/SiC system demonstrate that high-contrast active tuning of both modes can be realized via switching of the GST phase, achieving an exceptional tuning figure of merit of up to $7.7$. We have presented a simple theoretical model for the waveguide mode, revealing the similiarity of the SPhP and waveguide mode dispersions in the reststrahlen region of the substrate material. Furthermore, by analyzing the electric field distributions across the system interfaces, we have shown that the waveguide mode can feature spatial confinement and field enhancement factors comparable to those of a SPhP, while expanding the allowed frequency range beyond the reststrahlen band. Thus, in addition to the polarization-limited surface polaritons, polariton-like waveguide modes supported by phase-change materials on polar crystal substrates provide a promising complementary building block for actively tunable, low-loss, and omni-polarized nanophotonic applications, such as in-plane metasurfaces or polariton lenses\cite{Folland2018,Chaudhary2019}.

\section{Methods}
\subsection{Experimental}
The \nmetr{144} thick film of the phase-change material $\text{Ge}_3\text{Sb}_2\text{Te}_6$ was deposited onto the SiC substrate by direct current magnetron sputtering with a background pressure of about $2\times 10^{-6}$ mbar and 20 sccm Ar flow in constant power mode (20 W) using stoichiometric targets of $99.99\%$ purity.

The experimental setup is illustrated in Fig. \ref{fig2}a and has been described before\cite{Passler2017,Passler2018}. In short, the Otto geometry is implemented employing three motorized actuators (Newport TRA12PPD) to control the relative position of the coupling prism (KRS5, 25 mm high, 25 mm wide, angles of \dg{30}, \dg{30} and \dg{120}, with the \dg{120} edge cut-off, Korth Kristalle GmbH) with respect to the sample. The motors push the prism against springs away from the sample and thus enable continuous tuning of the air gap width $d_{\text{gap}}$. The \dg{120} prism edge is cut-off parallel to the backside in order to quantify $d_{\text{gap}}$ via whitelight interferometry\cite{Pufahl2018a}. The smallest gap width achieved was of $\sim \mumetr{3}$. In principle, gap widths below \mumetr{3} can be measured, but were not feasible for the GST/SiC sample probably due to a not perfectly flat sample surface. 

For each incident angle and polarization, the air gap of critical mode coupling $d_{\text{crit}}$\cite{Passler2017} was determined by optimizing the mode resonance depth in the reflectance signal, and all spectra were taken at $d_{\text{crit}}$. However, $d_{\text{crit}}$ inversely scales with the resonance quality, because large losses into the sample require equally large radiative losses that are provided by the proximate coupling prism, thus leading to very small $d_{\text{crit}}$. This can lead to $d_{\text{crit}}$ being smaller than the experimentally achieved minimum gap of $\sim\mumetr{3}$, resulting in low resonance depths due to undercoupling\cite{Pufahl2018a}. This is, indeed, the reason why the waveguide mode resonance in the lossy c-GST maximally drops by $30 \%$ in our experiments (see Fig. \ref{fig2}a and \ref{fig3}), whereas all other resonances achieve $80\%$.

Furthermore, we note that the in-plane electric field distributions of SPhPs and waveguide modes in Fig. \ref{fig1}a-d, analogeous to Fig. \ref{fig5}a-f, were calculated for excitation via a KRS5 coupling prism in the Otto geometry at an incident angle of $\theta=\dg{30}$. For all four systems, the coupling gap was fixed to the critical coupling gap of the SPhP on bare SiC ($d_{\text{crit}}=\mumetr{5.3}$). This asymmetric excitation scheme is the reason for the asymmetric field distribution of the air-bound waveguide mode in Fig. \ref{fig1}d.

The mid-infrared FEL we employ as excitation source was operated at an electron energy of $31$ MeV, yielding a tuning range of the FEL outout wavelength between $\sim 7-\mumetr{18}$ ($\sim 1400-\wavenumber{550}$, covering the spectral range of the SiC reststrahlen band) by moving the motorized undulator gap. The micropulse repetition rate was set $1$ GHz, and the macropules had a duration of \musec{10} at a repetition rate of $10$ Hz, yielding ps-pulses with a typical FWHM$<\wavenumber{5}$ ($\sim 0.3 \%$). The macropulse and micropulse energies are $\sim 30$ mJ and \muJ{10}, respectively. However, the incident FEL beam was attenuated by -18 dB in order to avoid irreversible damage of the GST film and unintentional crystallization. Further details on the FEL are given elsewhere\cite{Schollkopf2015}.

The reflectance signal is detected with a pyroelectric sensor (Eltec  420M7). A second sensor measures the power reflected off a thin KBr plate placed prior to the setup in the beam. This reference signal is used to normalize all reflectance signals and is measured on a single shot basis in order to minimize the impact of shot-to-shot fluctuations of the FEL. The spectral response function of the setup is determined by taking a reflectance spectrum at large gap widths $d_{\text{gap}}\approx \mumetr{40}$, where the beam is totally reflected at the prism backside across the full spectral range. References are measured for each incident angle and polarization individually, and all reflectance spectra are normalized to their respective response function.

\subsection{Global Fitting Procedure}
The spectra in Fig. \ref{fig2}b, the dispersion curves in Fig. \ref{fig1}e,f, \ref{fig3} and \ref{fig6}, and the electric field distributions in Fig. \ref{fig5} were all calculated with a $4 \times 4$ transfer matrix formalism\cite{Passler2017a} that accounts for the anisoptropy of the c-cut hexagonal 4H-SiC substrate. For the calculations, most of the material parameters and system variables were set to values from literature or in advance determined values (including the phonon frequencies $\omega_{TO\text{,o}}^{\text{SiC}}=\wavenumber{797}$, $\omega_{TO\text{,e}}^{\text{SiC}}=\wavenumber{788}$, $\omega_{LO\text{,o}}^{\text{SiC}}=\wavenumber{970}$, $\omega_{LO\text{,e}}^{\text{SiC}}=\wavenumber{964}$ where o stands for ordinary and e for the extraordinary crystal direction\cite{Engelbrecht1993}, the damping $\gamma_{\text{SiC}}=\wavenumber{3.75}$, $\varepsilon_{\inf}^{\text{SiC}}=6.5$\cite{Neuner2009}, the spectral width of the FEL $\Delta\omega=\wavenumber{5.56}$, the angular width of the excitation beam $\Delta\theta=\dg{0.24}$ and all nine incident angles $\theta=26-\dg{34}$). The remaining parameters, on the other hand, were extracted from a global fitting procedure and comprise the a-GST film thickness $d_{\text{a-GST}}=\nmetr{144}$ (the c-GST film thickness was calculated by a constant factor, $d_{\text{c-GST}}=0.92 \: d_{\text{a-GST}}=\nmetr{132}$\cite{Njoroge2002,Michel2016a}), the real and imaginary part of the permittivity of c-GST ($\varepsilon_{\text{c-GST}}=31.4+9 i$), the real permittivity of a-GST ($\varepsilon_{\text{a-GST}}=13.8 + 0 i$), the gap widths $d_{\text{gap}}$ and a multiplication factor $a$ to account for slow FEL power drifts ($0.96<a<1$). The data sets of all four system configurations (p-polarization and s-polarization for a-GST and c-GST, respectively), each containing nine spectra at different incident angles, were fitted globally, where $d_{\text{a-GST}}$, $\varepsilon_{\text{a-GST}}$, and $\varepsilon_{\text{c-GST}}$ were global fitting parameters, while the factors $a$ and the gap widths $d_{\text{gap}}$ were adjusted individually for each spectrum. 

\begin{acknowledgement}
We thank Wieland Sch\"ollkopf and Sandy Gewinner for operating the FEL, and Martin Wolf and the Max Planck Society for supporting this work. We also thank Jialiang Gao for sputtering the GST film. This work was supported by the German Federal Ministry of Education and Research within the funding program Photonics Research Germany (contract number 13N14151) and the DFG (German Science Foundation) within the collaborative research center SFB 917 “Nanoswitches”. 
\end{acknowledgement}


\bibliography{GSTwaveguide}

\end{document}


\title{Supporting Information: \\ Surface Polariton-Like s-Polarized Waveguide Modes in Switchable Dielectric Thin-Films on Polar Crystals}

\author{Nikolai Christian Passler}
 \email{passler@fhi-berlin.mpg.de}
 \affiliation{Fritz Haber Institute of the Max Planck Society, Faradayweg 4-6, 14195 Berlin, Germany}
\author{Andreas He\ss ler}
 \affiliation{Institute of Physics (IA), RWTH Aachen University, 52056 Aachen, Germany}
\author{Matthias Wuttig}
 \affiliation{Institute of Physics (IA), RWTH Aachen University, 52056 Aachen, Germany}
\author{Thomas Taubner}
 \affiliation{Institute of Physics (IA), RWTH Aachen University, 52056 Aachen, Germany}
\author{Alexander Paarmann}
 \email{alexander.paarmann@fhi-berlin.mpg.de}
 \affiliation{Fritz Haber Institute of the Max Planck Society, Faradayweg 4-6, 14195 Berlin, Germany}

\date{\today}
\maketitle
\beginsupplement

\begin{figure}[t]
\includegraphics[width=.75\linewidth]{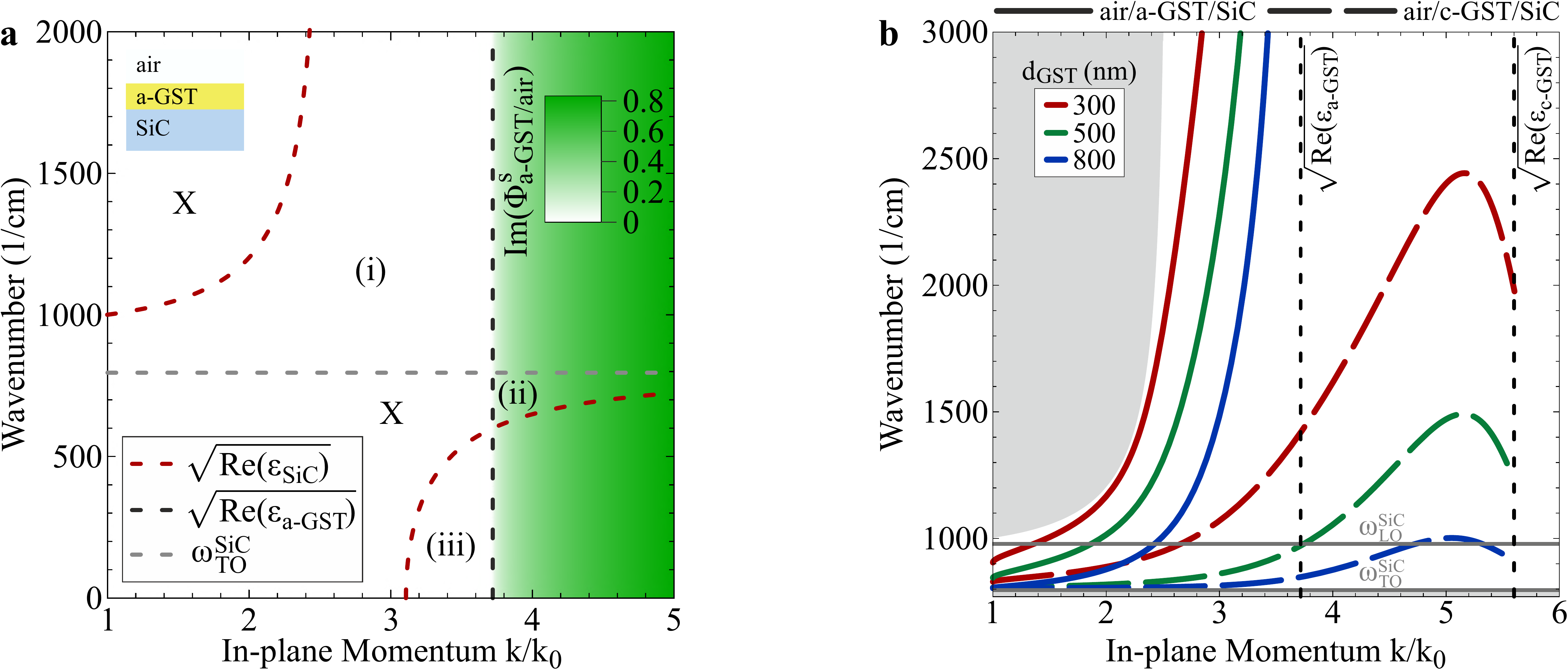}
  \caption{\textbf{Theoretical dispersion of the m=0 order s-polarized waveguide mode in a-GST and c-GST on SiC.} \textbf{a} Imaginary part of the phase difference $\text{Im}(\Phi_{\text{a-GST/air}}^s)$ upon reflection at the a-GST/air interface calculated with Eq. 2 from the main text\cite{Tien1970}. Where $\text{Im}(\Phi_{\text{a-GST/air}}^s)>0$, no waveguide modes can be supported, which is the case for $k/k_0>\sqrt{\text{Re}(\varepsilon_{\text{a-GST}})}$. Therefore, region (ii) is ruled out, even though the SiC/a-GST interface would allow a waveguide mode ($\text{Im}(\Phi_{\text{SiC/a-GST}}^s)=0$), see Fig. 4a of the main text. Regions marked with an X are excluded because there, $\text{Im}(\Phi_{\text{SiC/a-GST}}^s)>0$. Thus, the air/a-GST/SiC system only supports waveguide modes in regions (i) and (iii). \textbf{b} Comparison of the dispersion of an m=0 s-polarized waveguide mode in an air/a-GST/SiC (solid lines) and an air/c-GST/SiC system (dashed lines) for three different GST film thicknesses. The upper momentum boundary of the dispersions for both systems is given by the refractive index ($n=\sqrt{\varepsilon}$) of the respective GST phase (dashed black lines). The dispersions in a-GST, which has neglectable absorption in the shown frequency range ($\text{Im}(\varepsilon_{\text{a-GST}})=0$), approach the upper momentum boundary asymtotically, whereas the dispersions in c-GST experience a red-shift in proximity to the upper momentum boundary due to significant absorption ($\text{Im}(\varepsilon_{\text{c-GST}})=9$). For small in-plane momenta ($1<k/k_0<2$), however, which is the range studied in this work, the dispersions in the two GST phases are qualitatively the same, where in c-GST the modes are red-shifted compared to a-GST.}
  \label{sfig1}
\end{figure}

\begin{figure}[t]
\includegraphics[width=1\linewidth]{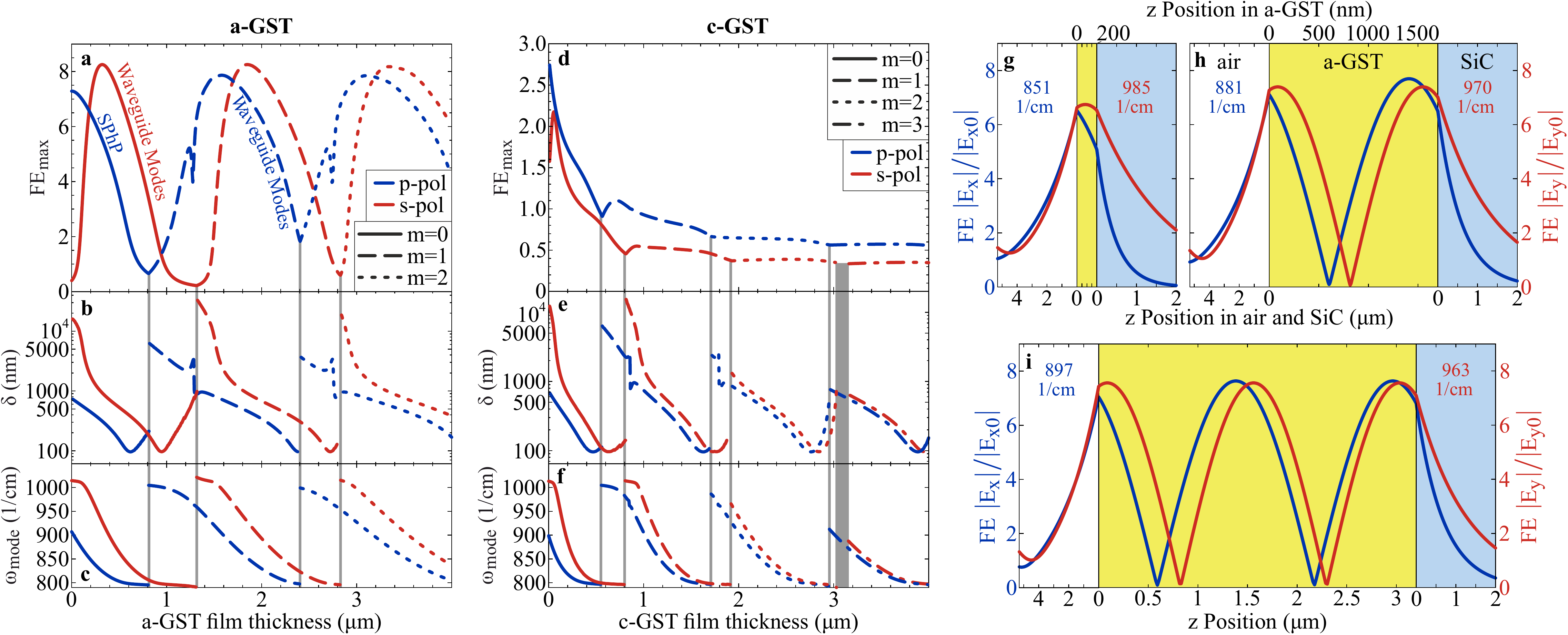}
  \caption{\textbf{Higher-order waveguide modes in a-GST and c-GST on SiC.} \linespread{1.15}\small{\textbf{a,d} Maximum field enhancement FE$_{\text{max}}$ for p- (blue curves) and s-polarization (red curves) as a function of a-GST and c-GST film thickness $d_{\text{GST}}$, respectively, at $\theta=\dg{30}$. The respective penetration depths $\delta$ into SiC are plotted in \textbf{b,e}, and \textbf{c,f} shows the corresponding mode frequencies $\omega_{\text{mode}}$. Above certain $d_{\text{GST}}$ (marked by gray lines where the mode frequencies jump), however, the first-order s- and p-polarized waveguide modes start being supported, appearing at the upper frequency boundary. Noteworthily, the SPhP mode propagating at a single boundary smoothly fits into the waveguide mode picture that dominates at larger $d_{\text{GST}}$, and could even be seen as a continuation of the p-polarized waveguide mode in the limit of vanishing film thicknesses. However, while all higher-order p-polarized ($m\!\geq \!1$) and all s-polarized waveguide modes are also supported in thin films on dielectric materials, the SPhP requires the negative permittivity of the SiC substrate. In summary, the reasons for a structure to support a SPhP or a waveguide mode are different, but otherwise SPhPs and waveguide modes can be treated very similarly. The penetration depth $\delta$ is a function of the SiC substrate permittivity, and is large for $\varepsilon > 0$ (above $\omega_{LO}^{\text{SiC}}=\wavenumber{970}$) while reaching a minimum of $\delta=\nmetr{100}$ for mode frequencies close to $\omega_{TO}^{\text{SiC}}=\wavenumber{797}$ (where $\varepsilon \ll 0$). For p-polarization, the field enhancement and the penetration depth experience a small, sharp feature where the mode crosses the region between the LO ordinary (o) and extraordinary (e) phonon frequencies $\omega_{LO\text{,o}}^{\text{SiC}}=\wavenumber{970}$ and $\omega_{LO\text{,e}}^{\text{SiC}}=\wavenumber{964}$, being an effect of the SiC c-cut uniaxial crystal anisotropy\cite{Razdolski2016}. In comparison to a-GST, the high absorption of c-GST leads to a significantly reduced field enhancement especially for the higher-order modes. Due to larger internal losses, the critical coupling gap of c-GST ($\mumetr{1.5}$) where the modes are excited optimally in the Otto geometry is smaller than for a-GST ($\mumetr{3.7}$). \textbf{g-i} Electric field profiles for p- and s-polarized incident light across an air/a-GST/SiC system for a-GST films of 200, 1700, and \nmetr{3200} film thickness, respectively, excited in the Otto geometry. For $d_{\text{a-GST}}=\nmetr{1700}$ (h), the $m=1$ waveguide modes for both polarizations are supported, featuring a node inside the a-GST film, where for $d_{\text{a-GST}}=\nmetr{3200}$ (i), the $m=2$ waveguide modes with two nodes inside the a-GST film can be observed.}}
  \label{sfig2}
\end{figure}

\clearpage

\bibliography{supplMat}